\documentclass[twocolumn,english,prl,amsmath,amssymb]{revtex4-1}
\usepackage[T1]{fontenc}
\usepackage{times}
\usepackage{graphicx}
\usepackage{amsmath}
\usepackage{amssymb}
\usepackage{natbib}
\usepackage{babel} % multi-language
\usepackage{bm} % bold math
\usepackage{xcolor} % text color
\usepackage{ulem} % strikethrough, e.g.
\usepackage{hyperref}
\begin{document}
\title{Nature of the Quantum Metal in a Two-Dimensional Crystalline Superconductor}

\author
{
A. W. Tsen,$^{1}$ B. Hunt,$^{1\dagger}$ Y. D. Kim,$^{2}$ Z. J. Yuan$^{3}$, S. Jia$^{3,4}$, R. J. Cava$^{5}$, J. Hone$^{2}$, P. Kim$^{6}$,\\
 C. R. Dean$^{1*}$ and A. N. Pasupathy$^{1*}$\\
\normalsize{$^{1}$\textit{Department of Physics, Columbia University, New York, NY 10027, USA}}\\
\normalsize{$^{2}$\textit{Department of Mechanical Engineering, Columbia University, New York, NY 10027, USA.}}\\
\normalsize{$^{3}$\textit{International Center for Quantum Materials, Peking University, Beijing 100871, China.}}\\
\normalsize{$^{4}$\textit{Collaborative Innovation Center of Quantum Matter, Beijing 100871, China.}}\\
\normalsize{$^{5}$\textit{Department of Chemistry, Princeton University, Princeton, NJ 08544, USA.}}\\
\normalsize{$^{6}$\textit{Department of Physics, Harvard University, Cambridge, MA 02138, USA.}}\\
\normalsize{$^\dagger$Current address: \textit{Department of Physics, Carnegie Mellon University, Pittsburgh, PA 15213, USA.}}\\
\normalsize{$^*$email: cd2478@columbia.edu; apn2108@columbia.edu  }
}

%\begin{abstract}

%\end{abstract}
\maketitle

% abstract w/ Abhay's changes
	\textbf{Two-dimensional (2D) materials are not expected to be metals at low temperature due to electron localization \cite{abrahams_scaling_1979}. Consistent with this, pioneering studies on thin films reported only superconducting and insulating ground states, with a direct transition between the two as a function of disorder or magnetic field \cite{goldman_superconductorinsulator_1998,haviland_onset_1989,fisher_quantum_1990,hebard_magnetic-field-tuned_1990,yazdani_superconducting-insulating_1995}. However, more recent works have revealed the presence of an intermediate quantum metallic state occupying a substantial region of the phase diagram \cite{ephron_observation_1996,christiansen_evidence_2002,qin_magnetically_2006,steiner_approach_2008} whose nature is intensely debated  \cite{das_existence_1999,das_bose_2001,dalidovich_phase_2002,phillips_elusive_2003,shimshoni_transport_1998,spivak_quantum_2001,galitski_vortices_2005}.  Here, we observe such a state in the disorder-free limit of a crystalline 2D superconductor, produced by mechanical co-lamination of NbSe$_2$ in inert atmosphere. Under a small perpendicular magnetic field, we induce a transition from superconductor to the quantum metal. We find a unique power law scaling with field in this phase, which is consistent with the Bose metal model where metallic behavior arises from strong phase fluctuations caused by the magnetic field \cite{das_existence_1999,das_bose_2001,dalidovich_phase_2002,phillips_elusive_2003}.
}

Global superconductivity emerges in a sample when conduction electrons form Cooper pairs and condense into a macroscopic, phase-coherent quantum state. In two dimensions, the phase coherence can be disrupted even at zero temperature by increasing disorder, either by degrading crystal quality or applying magnetic fields to create vortices \cite{goldman_superconductorinsulator_1998}. Granular or amorphous superconducting thin films, for which disorder levels can be controlled during growth, have thus provided an established platform for the study of quantum phase transitions in 2D superconductors. Within the conventional theoretical framework, increasing disorder or magnetic field perpendicular to a strongly disordered film at $T = 0$ induces a direct transition to an insulating state as the normal state sheet resistance approaches the pair quantum resistance $h/(2e)^2$ = 6.4 k$\Omega$ \cite{goldman_superconductorinsulator_1998,fisher_quantum_1990}. As film quality has improved over time, however, an intervening metallic phase with resistance much lower than the normal state resistance has been observed in several systems with generally less disorder \cite{ephron_observation_1996,christiansen_evidence_2002,qin_magnetically_2006,steiner_approach_2008}. Its origin is not well understood, and the various theoretical treatments can be generally divided between purely bosonic-based models, in which Cooper pairing persists in the metallic phase but phase coherence is lost \cite{phillips_elusive_2003,das_existence_1999,das_bose_2001,dalidovich_phase_2002}, and models that also incorporate other fermionic degrees of freedom \cite{shimshoni_transport_1998,spivak_quantum_2001,galitski_vortices_2005}.

	%%% FIGURE 1 %%%%%%%%%%%%%%%%
	\begin{figure}[bp!]
	\begin{center}
\includegraphics[scale=0.85]{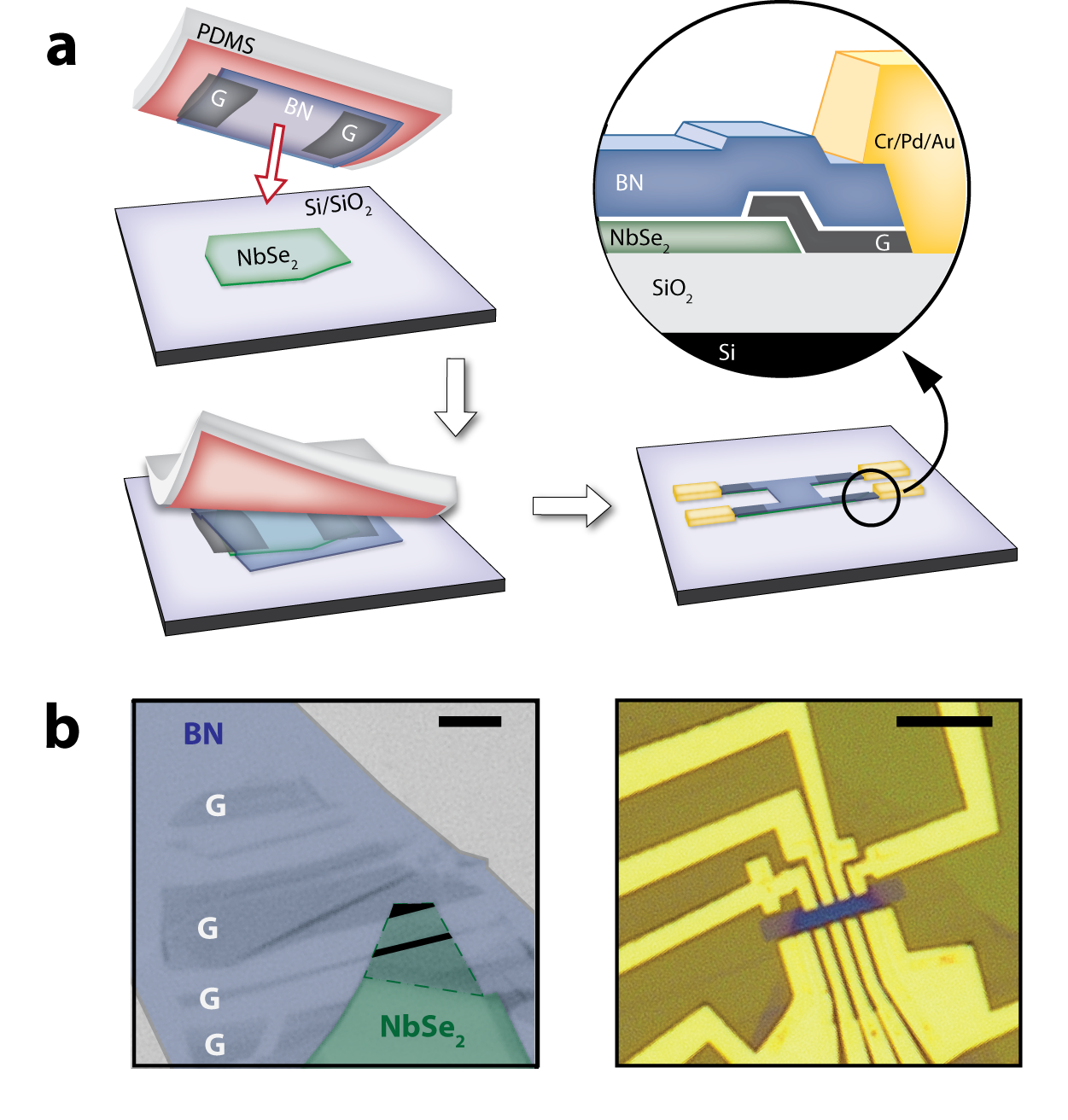}
\caption{\textbf{Environmentally controlled device fabrication}. \textbf{a)} Schematic of heterostructure assembly process. Boron nitride (BN)/graphite (G) on a polymer stamp (PDMS) is used to electrically contact and encapsulate NbSe$_2$ in inert atmosphere. The heterostructure is lithographically patterned and the edge of graphite is metallized with Cr/Pd/Au. \textbf{b)} Optical images of heterostructure before (left) and after device fabrication (right). In the (false-colored) left panel, the bilayer NbSe$_2$ is outlined in a dashed green line and the overlap between the graphite and bilayer NbSe$_2$ is shaded black.  Scale bar is 5 $\mu$m in both images.}
	\end{center}
\vspace{0mm}
\end{figure}	
%%%%%%%%%%%%%%%%%%%%%%%%%%%%%	

Recently, mechanical exfoliation has emerged as a technique to produce ultra-clean, crystalline 2D materials, with graphene being a well-known example \cite{geim_rise_2007}. Like amorphous films, the thickness of these samples can be easily controlled down to the level of individual atomic layers. In contrast to amorphous films, a 2D superconductor exfoliated from a layered, single crystal, can exist in the regime of minimal disorder, allowing for new insight into the nature of the vortex state in two dimensions. In this work, we realize such a system using a clean bilayer of NbSe$_2$, a well-known type-II superconductor with $T_c \sim 7.2$ K in bulk form \cite{trey_anisotropy_1973,soto_electric_2007}. Unique to this sample, the normal state sheet resistance is two orders of magnitude below $h/(2e)^2$ and insulating behavior is never observed. Instead, the intermediate metallic phase emerges in an exceptionally large region of the magnetic field-temperature phase diagram. Unlike the typical exponential behavior associated with the quantum tunneling of fermionic quasiparticles \cite{ephron_observation_1996,shimshoni_transport_1998}, we observe a new power-law scaling as a function of field at low temperature that is consistent with the Bose metal scenario of the metallic phase \cite{das_existence_1999,das_bose_2001,dalidovich_phase_2002,phillips_elusive_2003}.

Initial studies on exfoliated NbSe$_2$ flakes do not observe a superconducting transition to a zero-resistance state in the atomically thin limit \cite{staley_electric_2009,el-bana_superconductivity_2013}. Recently, however, it has been shown the surfaces of metallic materials may oxidize, altering the electronic properties of thin samples \cite{tsen_structure_2015}. Exfoliation and encapsulation by a protective layer in inert atmosphere is thus crucial for preserving the intrinsic properties of the 2D material \cite{tsen_structure_2015,cao_quality_2015}. We achieve this using a mechanical transfer setup installed inside a nitrogen-filled glove box (see Methods). In short, within the glove box, an exfoliated NbSe$_2$ flake is electrically contacted by graphite (G), and the entire device is protected by an insulating layer of hexagonal boron nitride (BN). The graphite leads are then contacted using an edge-metallization technique in the ambient environment \cite{wang_one-dimensional_2013,cui_multi-terminal_2015}. A schematic depicting the assembly and fabrication process is shown in Figure 1a and optical images of the heterostructure are shown in Figure 1b before (left) and after device fabrication.

%%%%%%%%%%%%%%%%%% FIGURE 2 %%%%%%%%%%%%%%%%%%%%%%%%%%%
\begin{figure}[htp!]
\begin{center}
\includegraphics[scale=0.95]{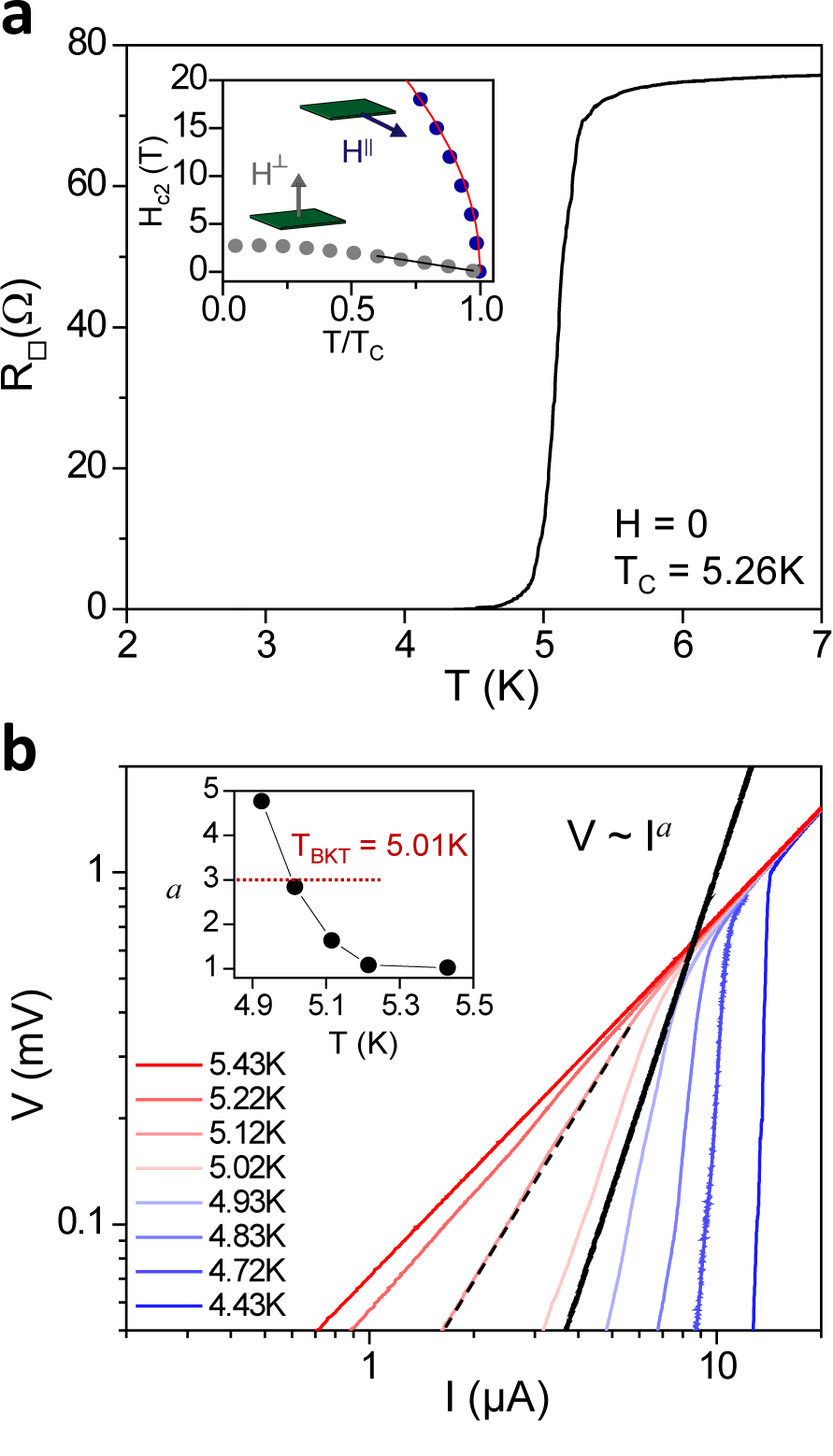}
\caption{\textbf{Characterization of bilayer NbSe$_2$ device. a)} Sheet resistance with temperature shows superconducting transition at $T_c$ = 5.26 K. Temperature-dependent critical magnetic fields parallel and perpendicular to the layers is shown in the inset. Black line is linear fit to $H_{c2}^{\perp}\propto 1-T/T_c$ at high temperatures. Red line is fit to $H_{c2}^{||}\propto \sqrt{1-T/T_c}$, the Tinkham formula for 2D samples \cite{tinkham_introduction_1996}. \textbf{b)} Voltage-current behavior at different temperatures. Inset shows exponent $a$ vs. $T$ extracted from power law fitting $V \sim I^a$ near the normal state transition. $a$ = 3 at the BKT temperature 5.01K.}
\end{center}
\end{figure}
\vspace{0mm}
%%%%%%%%%%%%%%%%%%%%%%%%%%%%%%%%%%%%%%%%%%%%%%%%%%%%%%%

%%%%%%%%%%%%%%%%% FIGURE 3 %%%%%%%%%%%%%%%%%%%%%%%%%%%
	\begin{figure*}[ht!]
	\begin{center}
\includegraphics[scale=0.6]{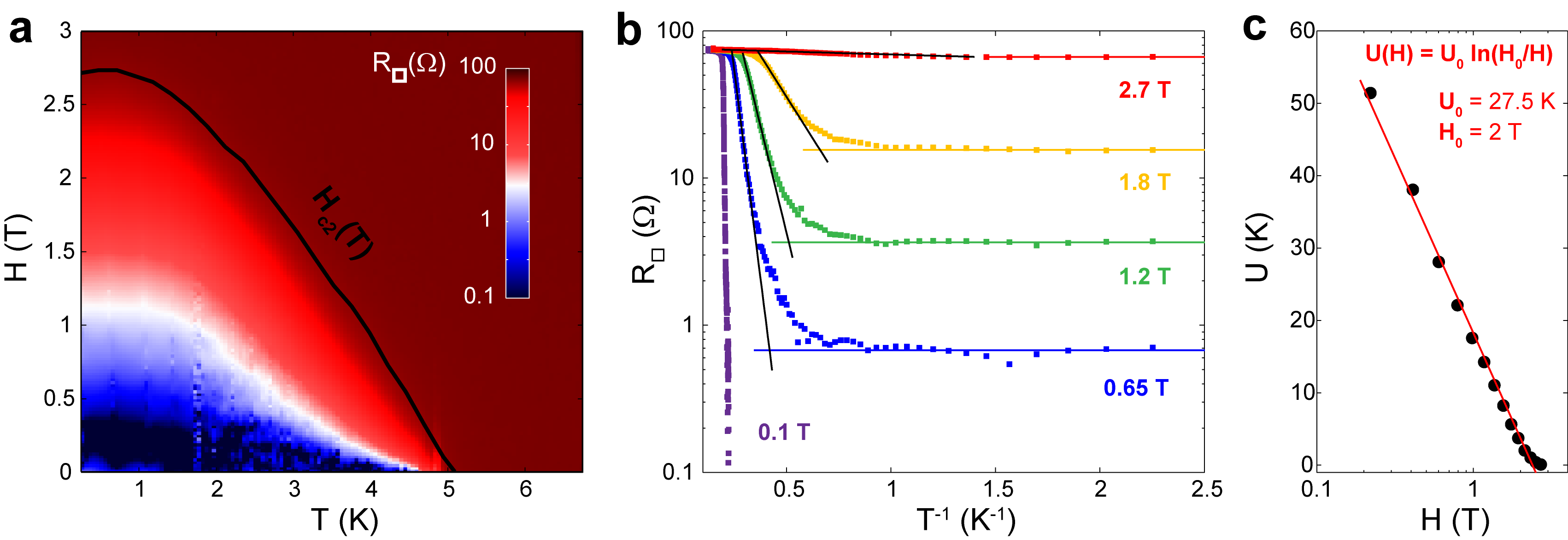}
\caption{\textbf{Magnetic field tuned phase transitions in 2D NbSe$_2$. a)} 2D color plot of sheet resistance vs. temperature and perpendicular magnetic field. \textbf{b)} Arrhenius plot of resistance for several magnetic fields shows thermally activated regime (black lines) and saturation at low temperatures (colored lines). \textbf{c)} Energy barrier vs. magnetic field extracted from linear fit to activated region. Solid red line is empirical fit to formula in inset.}
\end{center}
\end{figure*}
\vspace{0mm}
%%%%%%%%%%%%%%%%%%%%%%%%%%%%%%%%%%%%%%%%%%%%%%%%%%%%%%%

In the main panel of Figure 2a, we show four-terminal sheet resistance as a function of temperature for a particular NbSe$_2$ device prepared using the method described above. The NbSe$_2$ thickness is 1.5 nm as determined by an atomic force microscope (AFM) outside the glove box after BN/graphite transfer, suggesting it consists of only two atomic layers. The resistance in the normal state is $R_N$ = 75 $\Omega$. This corresponds to a residual resistivity that is 10 times larger than that of bulk crystals \cite{soto_electric_2007}, yet the sheet resistance is still an order of magnitude less than that of the most conductive amorphous superconducting films \cite{steiner_approach_2008}. We observe a clear superconducting transition to a zero-resistance state measured to the limit of our instrument resolution. The critical temperature is $T_c = 5.26$ K, as defined by where the resistance is 90\% of the normal state value, which is slightly reduced from the transition temperature of bulk samples (7.2K) \cite{trey_anisotropy_1973,soto_electric_2007}. 

In order to further characterize the quality and dimensionality of our sample, we have measured the temperature- dependent critical fields, defined by $R(H_{c2},T)=0.9R_N$, for field configuration both perpendicular and parallel to the layers, and the results are plotted in the inset of Figure 2a. Close to $T_c$, we expect $H_{c2}^{\perp}=(\Phi_0/2\pi\xi_0^2)(1-T/T_c)$, where $\Phi_0=h/2e$ is the flux quantum and $\xi_0$ is the in-plane coherence length at zero temperature. A linear fit shown by the black line yields $\xi_0$ = 8.9 nm, similar to the bulk value \cite{trey_anisotropy_1973,soto_electric_2007}.  From the normal state resistance and carrier concentration as determined by Hall measurements, we estimate the electron mean free path to be $\ell =$ 17 nm, smaller than in bulk crystals \cite{soto_electric_2007}, but nearly twice the coherence length, confirming that our device is in the pure superconductor regime \cite{tinkham_introduction_1996}.  In contrast, evaporated films are generally characterized as  ``dirty'' superconductors with $\xi_0 \gg \ell$.  The critical parallel field does not exhibit linear temperature dependence expected for anisotropic 3D superconductors \cite{tinkham_introduction_1996}. Instead, a 2D superconductor with thickness $d<\xi$ obeys $H_{c2}^{||}=(\sqrt{12}\Phi_0/2\pi\xi_0d)\sqrt{1-T/T_c}$ \cite{tinkham_introduction_1996}. The red curve in the inset of Figure 2a shows a best fit to this expression, from which we extract $d$ = 3.4 nm, which is slightly larger than the thickness as determined by AFM (1.5 nm).  However, this fitting has been previously found to overestimate the true sample thickness in sufficiently thin systems \cite{kim_intrinsic_2012}.  

The superconducting phase transition in a 2D material with $d<\xi_0$  is understood to be of the Berezinskii-Kosterlitz-Thouless (BKT) type \cite{halperin_resistive_1979}. In this scenario, the low-temperature, zero-resistance phase consists of bound vortex-antivortex pairs created by thermal fluctuations. Upon heating, the pairs dissociate and may move, inducing dissipation. The BKT temperature defines the vortex unbinding transition and can be determined using current-voltage measurements as a function of temperature $T$, as shown in the main panel of Figure 2b. Current excites free-moving vortices, causing a nonlinear voltage dependence: $V \sim I^{a(T)}$. At $T_{BKT}$, a 2D superconductor obeys the universal scaling relation, $V \sim I^3$ (solid line in Fig. 2b) \cite{halperin_resistive_1979,eley_approaching_2012}. In the inset, we plot $a$ vs. $T$, as determined by the slope of the different $V-I$ traces in log-log scale. An example guide-to-eye fit is marked by the dashed line in the main panel. We determine $T_{BKT}$ = 5.01 K from where $a = 3$ interpolates, only slightly less than $T_c$ as defined above. This is consistent with the behavior of systems with normal state resistance much less than $h/(2e)^2$, where $T_{BKT}$ is expected to be very close to the mean-field transition temperature \cite{beasley_possibility_1979}.

The measurements performed above confirm that our device exhibits the characteristics of a true 2D superconductor. The extracted material parameters together with the low normal state sheet resistance further places the high sample quality within a previously unexplored regime. We next turn to the dependence of resistance on perpendicular magnetic field as the effect of vortices can now be cleanly separated from the low static disorder present in the sample. Shown in Figure 3a is a 2D color map of the four-terminal sheet resistance in the same device as a function of both temperature and magnetic field applied perpendicular to the layers. For clarity, temperature traces for different field levels are shown in an Arrhenius plot in Figure 3b. For fields larger than $H_{c2}^{\perp} \sim 3$ T, the sample is in the normal state for all temperatures. Previous works on strongly disordered films observed a field-tuned transition to an insulating state of bosons \cite{hebard_magnetic-field-tuned_1990,yazdani_superconducting-insulating_1995}; however, one might not expect insulating behavior for finite samples of a highly conductive 2D system in the disorder-free limit \cite{steiner_approach_2008}. We have applied perpendicular fields as large as 14.5 T and still see metallic behavior in the temperature dependence (see Supplementary Information, Figure S1 main panel).

As one lowers the field to just below $H_{c2}$, a resistance drop is observed upon cooling from the normal state. In this region of the $H-T$ phase diagram, the device exhibits activated behavior, as can been seen in the linear slope in the Arrhenius plot (black lines in Fig. 3b). Classically, dissipation in a superconductor in which resistance is less than the normal state value can be attributed to the motion of individual vortices (flux creep or flow) \cite{tinkham_introduction_1996}. In a clean 2D system, we expect the dominant energy barrier to flux motion to be that of vortex-antivortex dissociation \cite{feigelman_pinning_1990}. Flux resistance then becomes thermally activated when the temperature is comparable to the barrier energy. We have determined the activation energy from the linear portion of the Arrhenius plot in Figure 3b for different magnetic fields, and the result is plotted in Figure 3c. The functional form is expected to be $U(H) = U_0 \ln(H_0/H)$, where $U_0=\Phi_0^2 d /(256 \pi^3 \lambda^2)$, the vortex-antivortex binding energy, and $H_0 \sim H_{c2}$ \cite{ephron_observation_1996,feigelman_pinning_1990}. A fit to this form yields $U_0$ = 27.5 K and $H_0$ = 2 T for our device. Assuming the magnetic penetration depth $\lambda$ is similar to the bulk value (230 nm) \cite{trey_anisotropy_1973}, we estimate $U_0 \sim$ 14 K, on the order of the experimental value. 

%%%%%%% FIGURE 4 %%%%%%%%%%%%%%%%%%%%%%%
	\begin{figure*}[ht!]
	\begin{center}
\includegraphics[scale=0.9]{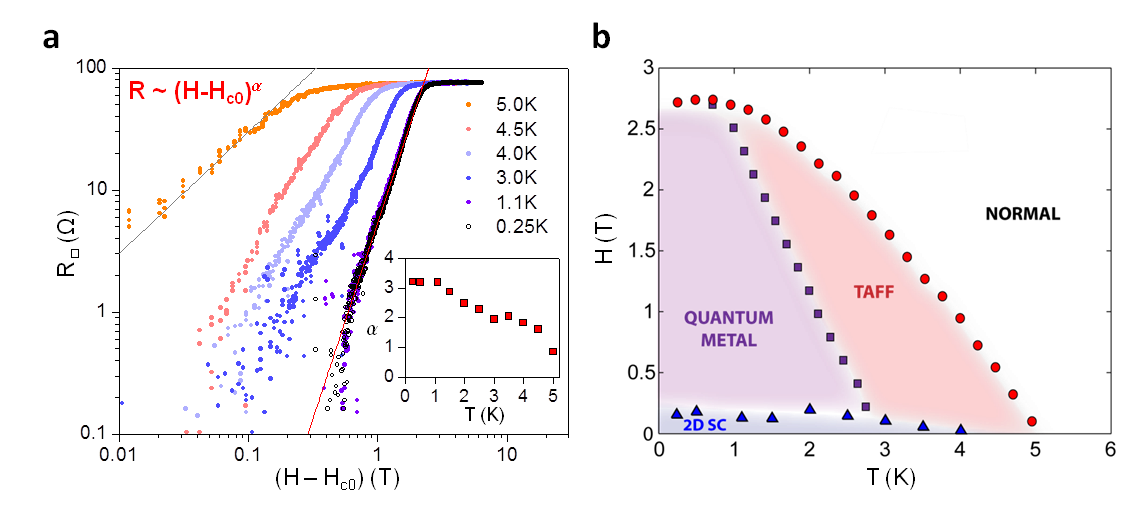}
\caption{\textbf{Emergence of the quantum metal. a) }Magnetoresistance above the superconducting transition for different temperatures. The data scale to a power-law $R\sim(H-H_{c0})^{\alpha(T)}$ and collapse onto a single curve in the quantum metallic phase below 1K. $\alpha$ vs. $T$ is plotted in inset. \textbf{b)} Full $H-T$ phase diagram of bilayer NbSe$_2$ device. The red circles are the locus of points where $R(H,T) = 0.9R_{N}$.  The purple squares, dividing the quantum metal from the thermally-assisted flux flow (TAFF) regime, show the transition from activated behavior $R\sim\exp(\Delta(H)/T)$ to temperature-independent $R=R(H)$, i.e. the intersection of the black and colored lines in Fig. 3(b).  The blue triangles denote the boundary of the superconducting phase $H_{c0}(T)$, the field at which resistance increases above the noise floor value. This criterion is determined by when hysteresis vanishes in $V-I$ measurements (see Fig. S4).}
	\end{center}
\vspace{0mm}
\end{figure*}
%%%%%%%%%%%%%%%%%%%%%%%%%%%%%%%%%%%%%%%

At lower temperatures, the resistance saturates to a level dependent on magnetic field (see colored lines in Fig. 3b), hallmark of the quantum metallic state that is the main subject of this report. Qualitatively similar behavior has also been observed in amorphous MoGe and Ta films with much larger normal state sheet resistances \cite{ephron_observation_1996,qin_magnetically_2006}. This effect cannot be understood within a classical framework, in which we expect to recover superconductivity for $T \ll U$. In the past, various theories have been advanced to explain the origin of dissipation in this temperature regime for disordered films \cite{phillips_elusive_2003,das_existence_1999,das_bose_2001,dalidovich_phase_2002,shimshoni_transport_1998,spivak_quantum_2001,galitski_vortices_2005}. One can distinguish between the different models based on the magnetic field dependence of the saturated resistance at low temperatures.

Shimshoni\textit{ et al. }consider a disordered superconducting film as a percolating network of superconducting islands within an insulating matrix \cite{shimshoni_transport_1998}. Field-induced vortices that tunnel across thin superconducting constrictions give rise to resistance when coupled to a fermionic bath. The resistance depends on field as  $R=h/(4e^2 ) \exp[C \pi/2 ((\hbar / e^2 )/R_N )((H-H_{c2})/H_{c2} )]$, where $R_N$ is the normal state resistance and $C$ is a dimensionless constant of order unity. This expression finds good agreement with the measurements of Ephron et al. on amorphous MoGe films, in which $R_N \sim $1 k$\Omega$ \cite{ephron_observation_1996}, but evidently provides a poor fit to our data (see Supplementary Information, Figure S2). Furthermore, since our device is over an order of magnitude more conductive, the closest fitting requires $C = 0.14$, an unphysically small value.

Galitski\textit{ et al. }consider a ``vortex metal'' phase where field-induced vortices interact with electrically neutral spinons \cite{galitski_vortices_2005}. The theory accounts for the large peak in low-temperature magnetoresistance observed in disordered InOx films before saturation to $R_N$ at higher fields \cite{steiner_approach_2008}. In our device, however, we observe a monotonic dependence of resistance on field and see no peak structure in magnetoresistance up to 7 T (see Supplementary Information, Figure S1 inset).

Das and Doniach, as well as Dalidovich and Phillips, report that a 2D system of interacting bosons may form a gapless, nonsuperfluid state in the limit of zero temperature, a phase which they term a ``Bose metal'' \cite{das_existence_1999,das_bose_2001,dalidovich_phase_2002,phillips_elusive_2003}. They argue that the uncondensed Cooper pairs and vortices are responsible for the small resistance observed at small finite field \cite{das_bose_2001}. A magnetic field introduces gauge fluctuations, which disrupts phase coherence and causes dissipation, a quantum analog of that caused by thermal fluctuations in the BKT transition. In the Bose metal model, resistance on the metallic side of the field-tuned transition can be described by  $R \sim (H-H_{c0})^{2\nu}$, where $H_{c0}$ is the critical field of the superconductor to Bose metal transition and $\nu$ is the exponent of the superfluid correlation length, which diverges across the boundary as $ (H-H_{c0})^{-\nu}$.

In the main panel of Figure 4a, we show a log-log plot of $R$ vs $H-H_{c0}$ taken at several different temperatures. $H_{c0}$ is a small temperature dependent value which we determined using a method that shall be described below. The linear scaling observed here suggests a power-law dependence on field. We have fit the data to the expression  $R \sim (H-H_{c0})^a$ and the extracted exponent $a$ is plotted in the inset as a function of temperature. At high temperatures, but below $T_c$, resistance increases roughly linearly with field, as expected for unhindered flux flow: $R \sim R_N H/H_{c2}$ ($a$ = 1) \cite{tinkham_introduction_1996}. The gray line shows linear scaling as a guide-to-eye. As temperature is lowered, the field dependence becomes increasingly nonlinear and collapses onto a single curve below 1 K with $a \sim 3$. The red line is an empirical fit given by  $R [ \Omega ] = 5.44(H [ T ]-0.15)^{3.21}$, which shows excellent agreement with the data. This yields a critical exponent of $\nu$ = 1.61. While previous measurements on MoGe films at the lowest accessible temperatures observe mostly an exponential dependence of resistance with field \cite{ephron_observation_1996,mason_true_2001}, which is consistent with quantum tunneling of fermionized vortices \cite{shimshoni_transport_1998}, at very small fields for the MoGe film, however, the scaling obeys a power-law with unity exponent \cite{mason_true_2001}. The power-law scaling we observe in our sample consistently over the entire field range thus suggests that the fermionic tunneling undergoes a crossover to Bose metal behavior in the limit of vanishing static disorder. We also found similar field scaling in the quantum metallic state for additional ultrathin devices, whereas this phase is distinctly absent in the bulk crystal (see Supplementary Information, Figure S3). 

The critical field of the transition out of the true zero-resistance state $H_{c0}$ is difficult to determine directly from the linear resistance given the limited accuracy of our instruments. Recently, however, experiments by Qin et al. and Li et al. on disordered Ta films showed that $H_{c0}$ can be determined indirectly using current-voltage measurements \cite{qin_magnetically_2006,li_transport_2010}. Hysteresis is observed in the $V-I$ characteristics on the superconducting side, similar to that seen in underdamped Josephson junctions \cite{tinkham_introduction_1996}, which disappear at the onset of the metallic phase. We have performed $V-I$ measurements on the same bilayer device at $T$ = 0.5 K for increasing magnetic fields (see Supplementary Information, Figure S4 main panel). Clear hysteresis is seen between current sweep up and down for low fields. In the inset, we plot the current hysteresis $\Delta I$ for where the voltage jump occurs as a function of magnetic field. $\Delta I$ vanishes close to $H_{c0}$ = 0.175 T. This then allows us to identify the true superconducting phase as that for $H < H_{c0}$. We have repeated this measurement for several different temperatures in order to determine critical field as a function of temperature.

Figure 4b shows a full $H-T$ phase diagram for our device. The blue triangles mark $H_{c0}(T)$ and the red circles mark $H_{c2}^{\perp} (T)$. The boundary between quantum metal and flux flow (purple squares) is defined by the intersection of the fits to the activated resistance and saturated levels in Figure 3b. The normal phase extends up to at least 14.5 T without the appearance of an insulating phase, in contrast to previous works on strongly disordered films. The high sample quality is made possible by our facile device assembly technique in inert atmosphere, which demonstrates a new route for the production of 2D superconductors in the ultraclean limit.

\textit{Methods.}  

\textit{(1) Crystal synthesis.} Polycrystalline NbSe$_2$ was made by heating stoichiometric amount of Nb powder (99.5\%) and Se shots (99.999\%) in evacuated silica ampoules. Single crystals of NbSe$_2$ were grown by using a vapor transfer method. 400 mg of NbSe$_2$ powder and 80 mg of I$_2$ were sealed in a 23cm-long silica ampoule with a 1.13 cm$^2$ inner cross section. The charge was put in the hot zone of 850$^{\circ}$C and the sink is in the cold zone with 750$^{\circ}$C. After one week, all the polycrystals became single crystals, while most of the thick plates of single crystals were found in the hot zone.

\textit{(2) Device assembly and fabrication.} We have exfoliated ultrathin NbSe$_2$ flakes in a nitrogen-filled glove box containing less than 2 ppm oxygen. Separately, we prepared thin hexagonal boron nitride (hBN) on a polydimethylsiloxane (PDMS) stamp covered with polypropylene carbonate (PPC), which we use to ``pick up'' two closely spaced graphite flakes (separation $\sim 2 \mu$m). The hBN/graphite stack is then used to cover the NbSe$_2$ flake inside the glove box.  While graphite makes electrical contact to NbSe$_2$, hBN provides an insulating oxidation barrier. Subsequent lithography may then be used to define a four-terminal device, in which the gapped region between the graphite leads form the channel. The graphite is then electrically contacted using an edge-metallization technique \cite{wang_one-dimensional_2013,cui_multi-terminal_2015}. 

\textit{Acknowledgements.} We acknowledge helpful discussions with Z. Han, J.-D. Pillet, E. Shimsoni, O. Vafek, A. Kapitulnik, S. Kivelson, D. Xiao, and D. Gopalan.  We thank J. Shi, F. Zhao, D. Wang, and S. Chen for assistance with device fabrication.  This material is based upon work supported by the NSF MRSEC program through Columbia in the Center for Precision Assembly of Superstratic and Superatomic Solids (DMR-1420634). Salary support is provided by the NSF under grants NEB- 1124894 (A.W.T.) and DMR-1056527 (A.N.P.). Some measurements were performed at the National High Magnetic Field Laboratory, which is supported by the NSF Cooperative Agreement (DMR-0654118), the State of Florida and the Department of Energy. S.J. is supported by the National Basic Research Program of China (grants 2013CB921901 and 2014CB239302). R.J.C. is supported by the Department of Energy, Division of Basic Energy Sciences (grant DOE FG02-98ER45706). P.K. acknowledges support from the Army Research Office (grant W911NF-14-1-0638).   

\textit{Author contributions.} A.W.T, B.H., C.D., and A.N.P. conceived and designed the experiment. Z.J.Y., S.J., and R.J.C. synthesized the NbSe$_2$ crystals. A.W.T. fabricated the devices with assistance from Y.D.K..  A.W.T. and B.H. performed the transport measurements. A.W.T, B.H., C.D., and A.N.P. analyzed the data and wrote the paper.  The authors declare no competing financial interests.

%\bibliography{references}

%%%%%%%%%%%%%%%%%%    %%%%%%%%    %%%%%%%%%%%


%%%%%%%%%%%%%%%%%%%    %%%%%%%%%%%%    %%%%%%%%%5

%\clearpage
%\newpage
%\section*{Supplementary Information}
%Contents: Supplementary Figures S1, S2 and S3.

%\newpage
%\begin{center}
%\textbf{Supplementary Figures}
%\end{center}

\setcounter{figure}{0}

%%%%%%% FIGURE S1 %%%%%%%%%%%%%%%%%%%%%%%
	\begin{figure*}[bp]
	 \renewcommand{\thefigure}{S\arabic{figure}} % only needs "\setcounter..." command at beginning
\begin{center}
	\includegraphics[scale=0.9]{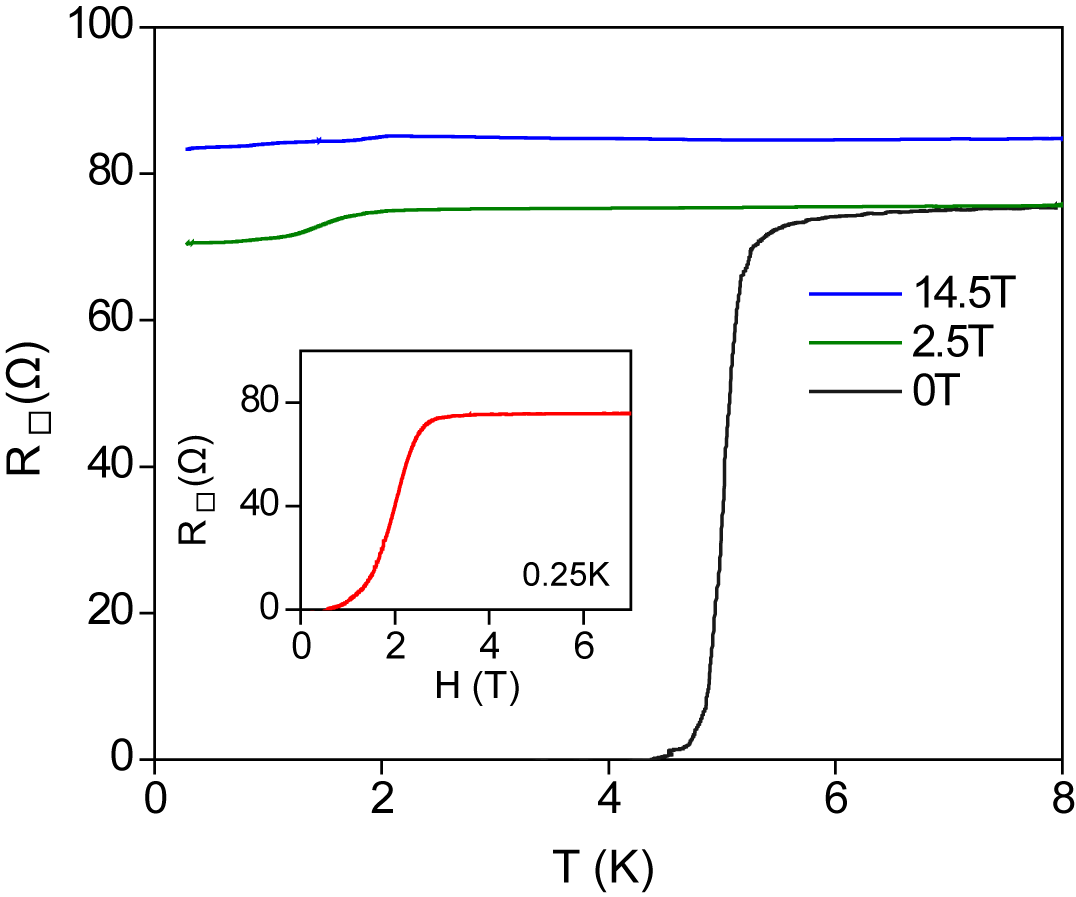}
\caption{\textbf{Behavior at higher magnetic fields. } Sheet resistance with temperature for several different perpendicular fields. No insulating behavior is observed up to 14.5T. Inset shows magnetoresistance for several different temperatures. No peak is observed up to 7T.}
	\end{center}
\vspace{0mm}
\end{figure*}
%%%%%%%%%%%%%%%%%%%%%%%%%%%%%%%%%%%%%%%

%%%%%%% FIGURE S2 %%%%%%%%%%%%%%%%%%%%%%%
	\begin{figure*}[bp!]
		 \renewcommand{\thefigure}{S\arabic{figure}} % only needs "\setcounter..." command at beginning
\begin{center}
	\includegraphics[scale=1]{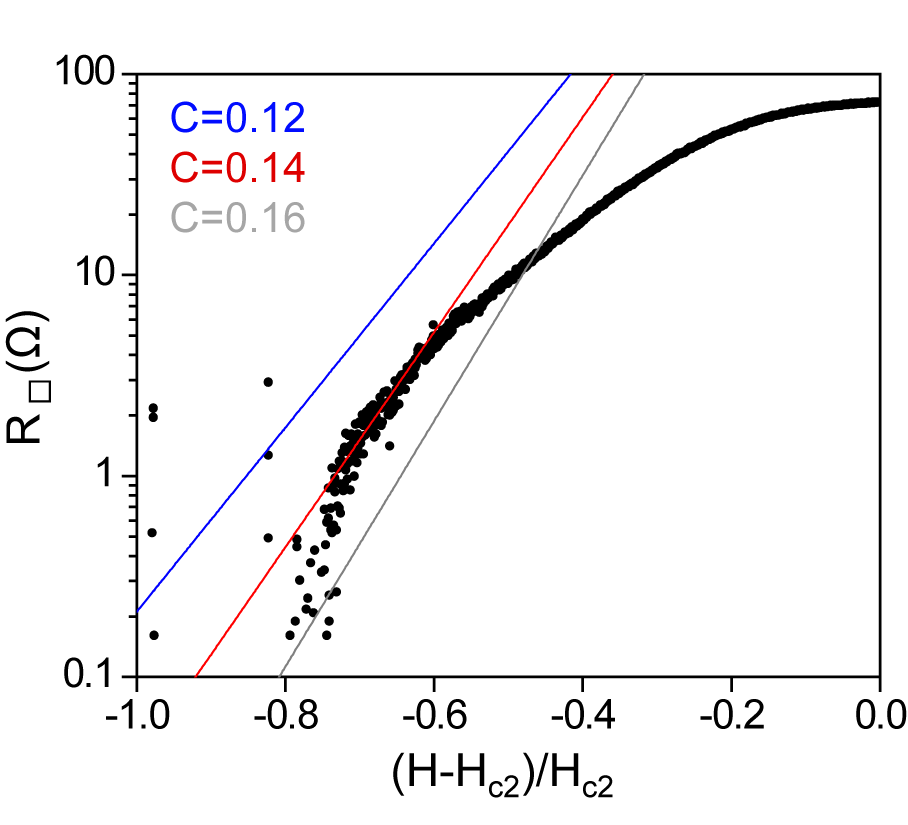}
\caption{\textbf{Vortex tunneling scenario.} Magnetoresistance at 0.25K fit to the formula in Shimshoni et al. describing quantum tunneling of vortices with several different values for constant $C$ \cite{shimshoni_transport_1998}. Best fitting ($C = 0.14$) deviates significantly at higher fields.}
	\end{center}
\vspace{0mm}
\end{figure*}
%%%%%%%%%%%%%%%%%%%%%%%%%%%%%%%%%%%%%%%

%%%%%%% FIGURE S3 %%%%%%%%%%%%%%%%%%%%%%%
	\begin{figure*}[bp!]
		 \renewcommand{\thefigure}{S\arabic{figure}} % only needs "\setcounter..." command at beginning
\begin{center}
	\includegraphics[scale=0.8]{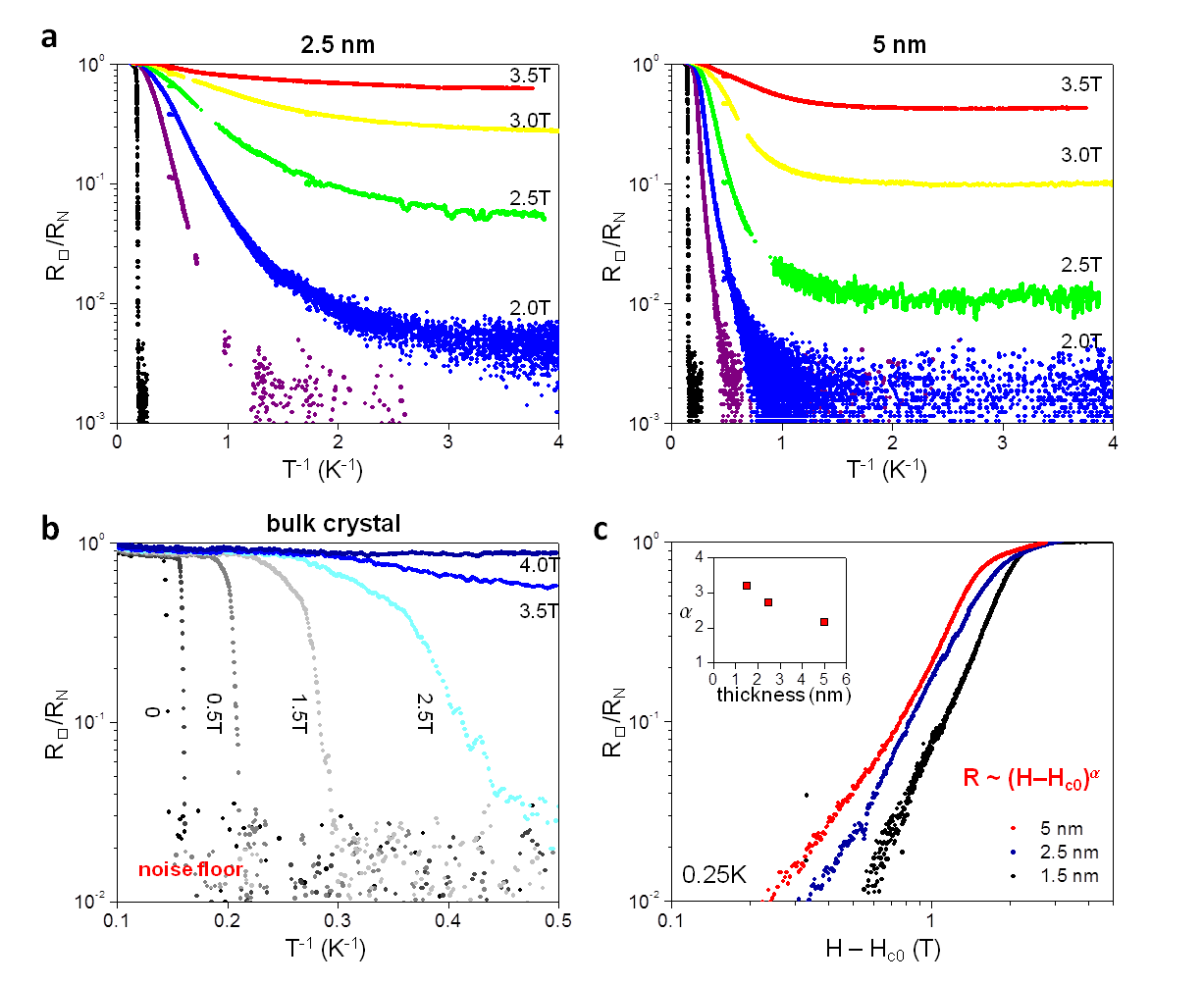}
\caption{\textbf{Thickness dependence of metallic behavior.} \textbf{a)} Arrhenius plot of normalized sheet resistance with different magnetic field levels for 2.5 nm (left) and 5 nm (right) thickness NbSe$_2$ devices.  Both show metallic behavior at low temperature and intermediate fields. \textbf{b)} Same plot for bulk crystal shows absence of the quantum metallic state. Transition temperature is reduced for increasing magnetic field. \textbf{c)} Magnetoresistance below the superconducting transition for different thin flake devices at 0.25K scaled to a power-law $R \sim (H - H_{c0})^\alpha$. $\alpha$ vs. thickness is plotted in inset.}
	\end{center}
\vspace{0mm}
\end{figure*}
%%%%%%%%%%%%%%%%%%%%%%%%%%%%%%%%%%%%%%%

%%%%%%% FIGURE S4 %%%%%%%%%%%%%%%%%%%%%%%
	\begin{figure*}[bp!]
		 \renewcommand{\thefigure}{S\arabic{figure}} % only needs "\setcounter..." command at beginning
\begin{center}
	\includegraphics[scale=1.0]{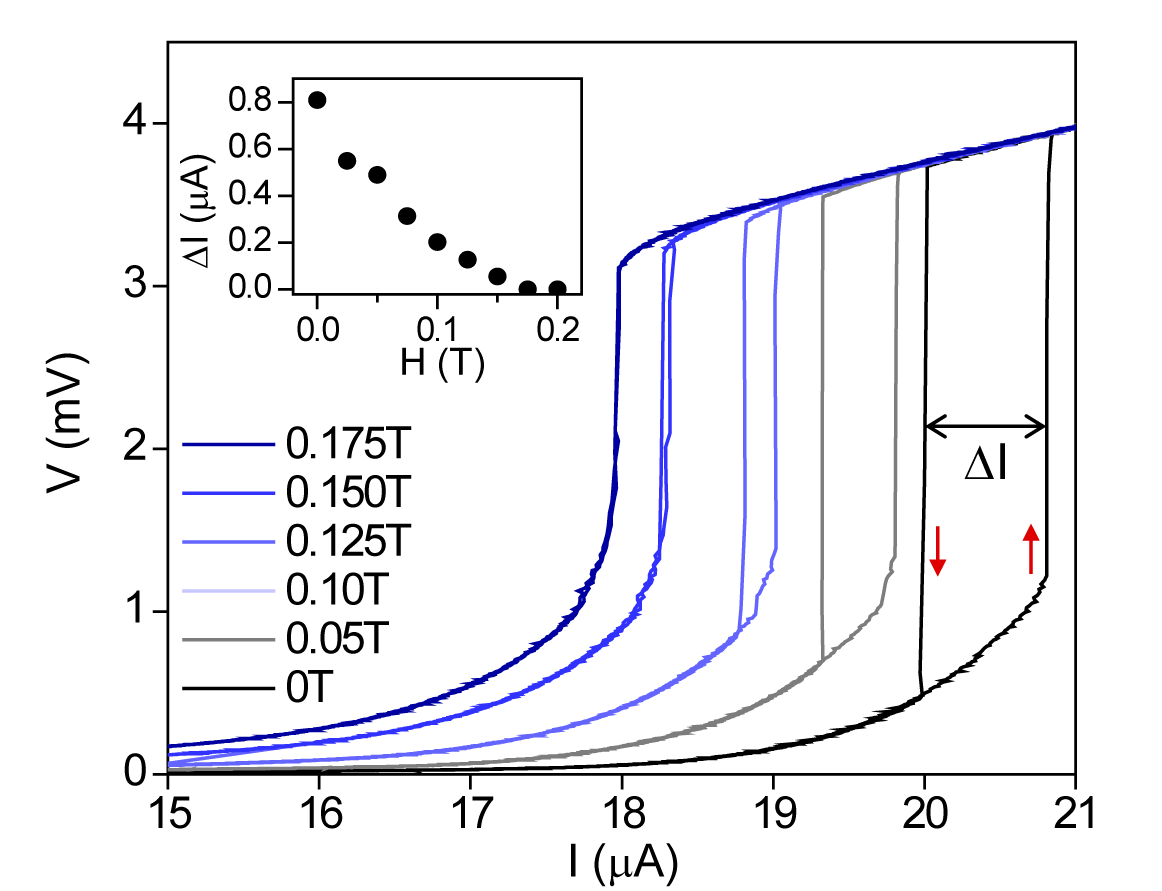}
\caption{\textbf{Field-tuned voltage-current characteristics.} $V-I$ traces at 0.5 K for several perpendicular field levels. Inset shows hysteresis between current sweep up and sweep down as a function of field. Hysteresis is observed in the superconducting phase, $H < H_{c0}$ = 0.175 T, but disappears for the metallic phase, $H > H_{c0}$. }
	\end{center}
\vspace{0mm}
\end{figure*}
%%%%%%%%%%%%%%%%%%%%%%%%%%%%%%%%%%%%%%%

 \end{document}